# Problems with *"GR without SR: A gravitational-domain description of first-order Doppler effects"* (gr-qc/9807084)

**Eric Baird**   *February 2001*

Einstein's goal of producing an advanced gravitational model that was independent of special relativity's "non-gravitational" derivations has arguably still not been achieved. The author produced a paper in 1998 outlining a possible method of attack on the problem based on shift equivalence principles ("Doppler mass shift"). Some problems with this paper have since come to light. We list some further developments, some papers not cited in gr-qc/9807084, some experimental results missed by the author, and identify a problem with the paper's approach to aberration effects.

## 1. INTRODUCTION

In April 1950, Einstein published an article in Scientific American stating his opinion that a full gravitational model should not depend on special relativity as a foundation [1] – while physics within flat spacetime is certainly very convenient [2], the simplifying assumption of inertial mass without gravitational mass is not a natural feature of more advanced gravitational models.

Paper gr-qc/9807084 by this author "GR without SR" [3] described a paradigm intended to avoid special relativity's "flat" geometry, with velocity-dependent curvature between relatively-moving masses acting as a regulating mechanism for local lightspeed constancy (with conventional Doppler shifts appearing as gravitational-domain effects). The most suprising aspects of this alternative model were **i)** that it seemed to allow a different "relativistic" shift equation to that used by special relativity, and **ii)** that this altered relationship did not seem to generate the usual incompatibilities between classical and quantum models ("black hole information paradox" [4]), and still seemed to be compatible with all current experimental data.

Since the first version of the paper was uploaded to the LANL archive, some problems have come to light, namely

1. the paper's accidental reversal of some aberration relationships and effects, and
2. the omission of published experimental evidence supporting special relativity's Doppler relationships.

In this supplementary paper we document these problems, and also list some additional work that has been produced or become known to the author since September 1998.

## 2. ABERRATION PROBLEMS

A basic derivation of relativistic aberration effects for shift equations **(1)**, **(2)** and **(3)** (as identified in [3] section 5.1) has shown that special relativity's aberration formula [5], with its forward-tilted rays is a "general" relativistic result [6], and does not depend on the assumption of flat spacetime or on special relativity's particular choice of Doppler equations.

The paper therefore seems to be in error when it relates a rearward-observed redshift on a receding object to an apparent rearward concentration of fieldlines in a receding object – while the author's calculated fieldline density-change ratios are probably correct, the actual relationship seems to be inversely proportional rather than directly proportional (it is easy to make this sort of accidental inversion when working with an "unexplored" model, e.g. Newton [7]).

The correct argument would seem to be the one presented elsewhere in the paper for the inverse relationship between the object's perception of environmental Doppler effects and of environmental fieldline density – the "moving" object sees more environmental mass ahead than behind, but the proposed Doppler mass shift effect acts to enhance the gravitational influence of aft (redshifted) masses and lessen the attraction of the forward (blueshifted) material – the Doppler effect acts *against* the calculated effect of aberration on fieldline density [8].

A recent piece by Carlip [9] has come to a broadly similar conclusion regarding the equal magnitude of displacement and aberration effects, with velocity components acting to make any supposed gravitational-field aberration effects undetectable.

http://arXiv.org/abs/physics/0102082

**physics/0102082**

### 3. MISSED EXPERIMENTAL RESULTS

#### 3.1   "Transverse" tests using non-transverse measurements

From "GR without SR" [3] section 9.1: "While it may seem improbable that we have not yet been able to verify that the SR shift prediction **(2)** is more accurate than the earlier equation **(1)**, the author has so far been unable to find any direct evidence favoring **(2)**"

In fact, a number of experiments that are listed as tests of special relativity's "transverse" redshift predictions *are* actually based on analysis of non-transverse data (e.g. Ives-Stilwell [10]). These experiments had been missed because of their usual classification as "transverse" tests [11].

At least two of the experiments listed in MacArthur's review piece [11], [10][12] do lend themselves to reanalysis, and do appear to give a very good match to the predictions of **(2)** rather than **(1)**, which seems to settle the issue in favour of special relativity. However, these experiments were carried out at a time when it seems to have been widely accepted that lab-transverse redshift effects could *only* be generated by **(2)** ("... no change in frequency." [13] "classically one would not expect a frequency shift from a source that moves by right angles." [14]), suggesting the possibility that an experimenter looking for either a null result or a Lorentz redshift might attribute an "inexplicable" double-strength redshift [15] to a combination of **(2)** and some additional redshifting effect in the apparatus (such as mirror recoil).

**Observations made at 90°$_{LAB}$**

|  | Eqn **(1)** | Eqn **(2)** | Eqn **(3)** |
|---|---|---|---|
| "bad" textbook predictions [13] | *null result* | Lorentz redshift | *null result* |
| Corrected predictions [15] | *null result* | Lorentz redshift | Double Lorentz redshift |

Since these experiments seem to be the only ones to date that give unequivocal support for special relativity's choice of shift equation, it would be helpful if similar future experiments [16] could be designed to test for possible agreement with **(1)**, as well as with **(2)** and **(3)** [17].

### 4. ADDITIONAL REFERENCES AND FURTHER WORK

#### 4.1   Black hole information paradox

This paradox (Susskind review article [4]) has also been discussed in detail by Preskill, Danielson and Schiffer [18][19], and does not appear to apply to a physics based on (1). Unruh's work on indirect radiation through signal horizons in non-SR models [20][21] described "sonic horizon" radiation as being an analogue of Hawking radiation. Visser has since presented this effect as a full Hawking radiation effect [22][23][24]. Visser's paper appeared in the LANL archive after research for "GR without SR" had been completed.

#### 4.2   Aberration and angle-dependent shifts

The angle-changes and angle-dependent frequency-shifts associated with hypothetical relativistic models based on **(1)**, **(2)** and **(3)** have now been calculated from general principles, without presupposing flat spacetime [6]. In this exercise, all three calculations generate the same aberration formula as special relativity [5], and generates the result that for any given laboratory angle, **(2)** predicts wavelengths that are Lorentz redshifted compared to **(3)**, but **(1)** predicts wavelengths that are *doubly* Lorentz-shifted compared to **(3)** [6][15]. Lorentz relationships are often presented as being unique to **(2)** (but *see* Visser [25]).

#### 4.3   E=mc$^2$

An exact derivation of the E=mc$^2$ result from **(1)** for a pair of plane-waves aligned with the moving object's path has now been given and discussed in [26]. The relationships given in [6] also let us apply this result to pairs of plane-waves emitted at any other angle.

#### 4.4   Transverse effects

The Lorentz-squared redshift predictions for a lab-transverse detector for **(1)** are general and apply to relativistic and non-relativistic calculations (*see:* Lodge's 1893 "spurious or apparent Doppler" prediction [27]). Special relativity's predictions are the root product of the predictions made for **(1)** and **(3)** not just for the non-transverse case [28], but also for any other angle defined in a given frame [29].

### 4.5    Time dilation without acceleration

Textbooks tend to suggest that muon pathlengths are only explainable using special relativity [30]. The mathematics indicates otherwise – for a muon created at the edge of the Earth's atmosphere with a given rest mass, rest frame lifetime and momentum, the "new" calculated penetration depth under special relativity is the same as the older Newtonian prediction [15].

Lab-transverse redshifts ("aberration redshifts") already feature in a range of models that do not include physical time-dilation effects [15].

### 4.6    Acceleration effects

Acceleration of an object towards the observer introduces non-linear behaviour that is not compatible with flat-space approximations [31] (see also "acceleration radiation" and Bekenstein/Hawking radiation under more standard theory). A full gravitational description of the "twins" problem can run into similar problems under general relativity [32][33].

## 5.    WORK BY OTHER AUTHORS

**Matt Visser** and **W.G. Unruh** have derived purely classical indirect radiation effects in transonic fluid flows, and identified these effects as Hawking radiation effects. This work would also seem to apply to indirect radiation through the r=2M surface in systems of physics based around shift equation **(1)**.

**Steve Carlip** has produced a study of the aberration-gravity problem, and related the lack of additional gravitational aberration effects to the existence of gravitational velocity components [9].

**S. Dinowitz** has described a model that sounds similar in concept to the Doppler mass shift idea [34].

**Wolfgang Rindler** has described how general relativity could (in theory) have been developed in the Nineteenth Century, independently of special relativity [35].

## 6.    CONCLUSIONS

The sort of model described in [3] calls on several obscure areas of physics theory that have not yet been fully explored, because of complicating non-linearities or incompatibilities with special relativity. Progress in at least some of these areas is now being stimulated by work on the black hole information paradox.

Experiments such as Ives/Stilwell indicate that Nature's shift laws do obey **(2)** rather than **(1)**, apparently invalidating this approach and making the existence of a possible non-SR solution to the information paradox irrelevant.

However, these experiments were probably designed to differentiate between the "null shift" and "redshift" predictions of **(2)** and **(3)**. If experimenters were not aware of the (largely undocumented) double Lorentz redshift predictions associated with **(1)**, it is still conceivable that further experiments, designed to differentiate between **(1)** and **(2)**, may still tip the balance of evidence towards the non-SR equation.

---

**Radial Doppler frequency-changes and ruler-changes**

$$freq'/freq = (c-v)/c \quad \ldots (1)$$

$$freq'/freq = \sqrt{(c-v)/(c+v)} \quad \ldots (2)$$

$$freq'/freq = c/(c+v) \quad \ldots (3)$$

( perceived ruler-lengths alter by the same ratio as perceived frequency )

**Lab-transverse Doppler frequency-changes and ruler-changes**

$$freq'/freq = 1 - v^2/c^2 \quad \ldots (t1)$$

$$freq'/freq = \sqrt{1 - v^2/c^2} \quad \ldots (t2)$$

$$freq'/freq = 1 \quad \ldots (t3)$$

( … these predictions apply to observations at a "90°" angle, defined in the laboratory observer's frame )